\documentclass[12pt,a4paper,fleqn]{article}
\usepackage{lineno}
\usepackage{graphicx}
\usepackage{caption}
\usepackage[margin=1in]{geometry}

\begin{document}

\begin{center}
    {\Large \bf Title: The prevalence of repeating fast radio bursts}
\end{center}
\vspace{1cm}
{\large Authors: Vikram Ravi$^{1,2,3,*}$.} \\

\vspace{0.3cm}

\noindent $^{1}$ Center for Astrophysics $|$ Harvard \& Smithsonian, 60 Garden Street, Cambridge MA 02138, USA. \\
$^{2}$ Cahill Center for Astronomy and Astrophysics, MC\,249-17, California Institute of Technology, Pasadena CA 91125, USA. \\
$^{3}$ E-mail: vikram@caltech.edu.

\clearpage


{\bf Fast radio bursts are extragalactic, sub-millisecond radio impulses of unknown origin$^{\mathbf{1,2}}$. Their dispersion measures, which quantify the observed frequency-dependent dispersive delays in terms of free-electron column densities, significantly exceed predictions from models$^{\mathbf{3}}$ of the Milky Way interstellar medium. The excess dispersions are likely accrued as fast radio bursts propagate through their host galaxies, gaseous galactic halos and the intergalactic medium$^{\mathbf{4,5}}$. Despite extensive follow-up observations of the published sample of 72 burst sources$^{\mathbf{6}}$, only two are observed to repeat$^{\mathbf{7,8}}$, and it is unknown whether or not the remainder are truly one-off events. Here I show that the volumetric occurrence rate of so far non-repeating fast radio bursts likely exceeds the rates of candidate cataclysmic progenitor events, and also likely exceeds the birth rates of candidate compact-object sources. This analysis is based on the high detection rate of bursts with low dispersion measures by the Canadian Hydrogen Intensity Mapping Experiment$^{\mathbf{9}}$. Within the existing suite of astrophysical scenarios for fast radio burst progenitors, I conclude that most observed cases originate from sources that emit several bursts over their lifetimes. }

Thirteen fast radio bursts (FRBs) were published by the Canadian Hydrogen Intensity Mapping Experiment (CHIME) collaboration, including one repeating source (FRB\,180814.J0422$+$73) that I exclude from my analysis$^{8,9}$ (see Methods). These events were detected during a pre-commissioning phase when the instrument was not operating with its full sensitivity and field of view. The survey was conducted over less than $7.82\times10^{-5}$\,sky-years, implying an all-sky FRB rate floor of 300\,day$^{-1}$ in the 400$-$800\,MHz CHIME frequency band. Despite the systematic uncertainties, this is an order of magnitude greater than the rate of bright FRBs detected with the Australian Square Kilometre Array Pathfinder (ASKAP)$^{10}$. Additionally, although the ASKAP FRBs typically have lower excess dispersion measures (DMs) than FRBs detected with the more sensitive Parkes telescope, CHIME has a more than ten times higher detection rate than ASKAP at low excess DMs (Figure~1). This motivated the present analysis of the volumetric occurrence rate of FRBs. 

The paucity of direct distance measurements for FRBs, based for example on observations of FRB host galaxies, has meant that FRB volumetric-rate estimates have relied on ascribing dominant fractions of the excess DMs to the intergalactic medium (IGM)$^{1,11}$. I define the extragalactic DM, ${\rm DM}_{\rm X}$, as the difference between FRB DMs and predictions from models$^{3}$ of the Milky Way interstellar medium ($\mathbf{\mathbf{\rm DM}_\mathbf{\rm MW}}$). If  FRB ${\rm DM}_{\rm X}$ values were entirely built up from a homogeneous IGM comprising all cosmic baryons, the ${\rm DM}_{\rm X}$ range of the CHIME sample of 79--979\,pc\,cm$^{-3}$ would correspond to a comoving distance range of $0.34 - 2.52$\,Gpc [4]. This assumption, together with a host-galaxy contribution to ${\rm DM}_{\rm X}$ of 100\,pc\,cm$^{-3}$, was previously applied to an early sample of four FRBs from the Parkes telescope to derive a volumetric rate of $2.4\times10^{4}$\,Gpc$^{-3}$\,yr$^{-1}$ [1,11]. Substantial, hard to quantify uncertainties affect this estimate, including the sample completeness at different distances, the host-galaxy ${\rm DM}_{\rm X}$ contributions and the effects of cosmological evolution in the FRB population. 

Here I derive a robust lower limit on the FRB volumetric rate using the CHIME sample of low-${\rm DM}_{\rm X}$ bursts. For each FRB, I consider the following components of ${\rm DM}_{\rm X}$:
\begin{equation}
{\rm DM}_{\rm X} = {\rm DM} - {\rm DM}_{\rm MW} = {\rm DM}_{\rm MWhalo} + {\rm DM}_{\rm IGM} + {\rm DM}_{\rm host},
\end{equation}
where ${\rm DM}_{\rm MW}$ is estimated from the NE2001 model for the Milky Way ionised interstellar medium$^{3}$, ${\rm DM}_{\rm MWhalo}$ arises in the Milky Way hot gaseous halo$^{5}$, ${\rm DM}_{\rm IGM}$ arises in the IGM$^{4}$, and ${\rm DM}_{\rm host}$ arises in FRB host galaxies. I adopt a relation between ${\rm DM}_{\rm IGM}$ and cosmological distance that incorporates the primordial helium fraction and an IGM baryon fraction of 0.84 [12], with a normally-distributed scatter of 10\,pc\,cm$^{-3}$ [4]. Recent observations of ultraviolet and X-ray quasar absorption lines associated with the Galactic halo imply the presence of a substantial gas mass at temperatures between $10^{4}-10^{7}$\,K, in agreement with observations of other galaxies and cosmological simulations, that contributes between $50-80$\,pc\,cm$^{-3}$ out to the Milky Way virial radius$^{5}$. I therefore assume that ${\rm DM}_{\rm MWhalo}$ can have any value between $50$\,pc\,cm$^{-3}$ and the minimum of $80$\,pc\,cm$^{-3}$ and $({\rm DM}_{\rm X}-{\rm DM}_{\rm host})$ with equal probability.  I perform the analysis below for different characteristic  values of ${\rm DM}_{\rm host}$. These features imply a probability distribution, $P(<{\rm DM}_{\rm X} \, | \,d)$, for an FRB to have an extragalactic DM that is less than the expectation given a $d$ (see Supplementary Figure 1).

A comparison of the ${\rm DM}_{\rm X}$ distributions of FRB samples from ASKAP and Parkes suggests that ASKAP is increasingly incomplete at higher values of ${\rm DM}_{\rm X}$ because of its insensitivity to the faint, high-${\rm DM}_{\rm X}$ FRBs observed by Parkes$^{10}$. This effect, together with a possible correlation between FRB temporal widths and ${\rm DM}_{\rm X}$ [2,9], and additional unknown systematics in the CHIME observations$^{9}$, strongly suggests that CHIME is also increasingly incomplete at higher values of ${\rm DM}_{\rm X}$. I mitigate this unknown incompleteness by deriving a lower limit on the FRB rate, $R$, within a limited volume bounded at $d=d_{\rm limit}$ that contains the lowest-${\rm DM}_{\rm X}$ CHIME FRBs. I choose a $d_{\rm limit}$ such that  $P(<{\rm DM}_{\rm X}\, | \, d)$ is essentially unity for the lowest-${\rm DM}_{\rm X}$  FRB (180729.J1316$+$55, ${\rm DM}_{\rm X}=78.610$\,pc\,cm$^{-3}$), and 0.95 for the FRB with the second lowest ${\rm DM}_{\rm X}$ (FRB\,180810.J1159$+$83, ${\rm DM}_{\rm X}=122.134$\,pc\,cm$^{-3}$); for no ${\rm DM}_{\rm host}$ contribution, $d_{\rm limit}=426$\,Mpc. I also consider the case where $P(<{\rm DM}_{\rm X}\, | \, d)$ is essentially unity for the two aforementioned FRBs, and 0.95 for the event with the third lowest ${\rm DM}_{\rm X}$ (FRB\,180814.J1554$+$74; ${\rm DM}_{\rm X}=197.32$\,pc\,cm$^{-3}$; $d_{\rm limit}=739$\,Mpc). When $P(<{\rm DM}_{\rm X}\, | \, d)\sim1$, the probability of having observed the FRB within a volume bounded by $d$ is likely to also be approximately unity. 

Assuming that the so far non-repeating FRBs are independent events associated with distinct sources, their observed occurrence within the bounded volume can be modelled as a Poisson process. Lower limits on $R$ can then be derived from the number of observed events using the Poisson probability mass function (see Methods)$^{13}$. For the two cases of $d_{\rm limit}$ described above, I assume that two and three FRBs were observed respectively, and calculate corresponding conservative 90\%-confidence lower limits on $R$. Direct statistical inference on $R$ cannot be performed without specifying a model for the unknown incompleteness of the CHIME observations. Besides this incompleteness, the lower limits quoted below are conservative because they do not account for possible angular beaming of FRBs and cosmological time-dilation.  

I evaluate $d_{\rm limit}$ and the 90\%-confidence lower limit on $R$ for different values of ${\rm DM}_{\rm host}$, and the results are shown in Figure~2. The lower limit on $R$ is less constraining for $d_{\rm limit}$ corresponding to three observed FRBs than for $d_{\rm limit}$  corresponding to two observed FRBs. The lower limits on $R$ decrease as $d_{\rm limit}$ is increased to include more CHIME FRBs, which may indicate increasing incompleteness at higher values of ${\rm DM}_{\rm X}$. In the remainder of the text, I quote lower limits on $R$ for $d_{\rm limit}$ corresponding to two observed FRBs. For no assumed ${\rm DM}_{\rm host}$, $R>2.1\times10^{4}$\,Gpc$^{-3}$\,yr$^{-1}$, and for ${\rm DM}_{\rm host}=50$\,pc\,cm$^{-3}$, $R>2.4\times10^{5}$\,Gpc$^{-3}$\,yr$^{-1}$. Despite these lower limits being conservative, they are comparable to the estimate of $R\sim2.4\times10^{4}$\,Gpc$^{-3}$\,yr$^{-1}$ from the initial sample of four Parkes FRBs where ${\rm DM}_{\rm host}=100$\,pc\,cm$^{-3}$ was assumed$^{1,11}$. This may be explained by the uncertainties noted above in the latter, such as incompleteness in the Parkes sample given the large $d_{\rm limit}$ that was adopted. I note that such a large ${\rm DM}_{\rm host}$ would be in tension with the lowest-${\rm DM}_{\rm X}$ CHIME FRBs. 

The FRB rates derived herein can be compared with predictions for non-repeating FRB progenitors, and with the estimated birth rates of FRB sources (see Figure~2). All astrophysical FRB progenitors (excluding those attributed to new physics) are expected to be contained within galaxies, in many cases in regions with atypically dense interstellar medium$^{14}$. Simulations of the ${\rm DM}_{\rm host}$ values corresponding to typical locations within inclination-averaged early-type and dwarf galaxies suggest  ${\rm DM}_{\rm host}=23$\,pc\,cm$^{-3}$, and ${\rm DM}_{\rm host}=35$\,pc\,cm$^{-3}$ is likely characteristic of typical locations within face on late-type galaxies$^{15,16}$. Substantially larger values of ${\rm DM}_{\rm host}$ of a few hundred pc\,cm$^{-3}$ were recently inferred for the ASKAP FRB sample based on a comparison between FRB DMs from ASKAP, Parkes, and CHIME, and a tentative correlation between the ASKAP FRB sky-locations and catalogs of nearby galaxies$^{17}$.

Models for non-repeating FRB progenitor events rely on the destruction of highly magnetised white dwarfs (WDs) or neutron stars (NSs). These events are, however, likely too rare to produce FRBs at the observed rate. For example, FRBs may be produced upon the merger of two WDs$^{18}$, the merger of two NSs$^{19}$, or the collapse of an NS to a black hole$^{20}$. Gravitational-wave observations have demonstrated that NS-NS mergers occur at less than a quarter of the FRB volumetric-rate limit$^{21}$. Binary WDs most likely trace stellar mass with a possible preference for galactic disks, making it unlikely that all WD-WD mergers are found in under-dense regions of their host galaxies$^{22}$, as is required for this FRB progenitor channel$^{23}$. Further, only NS-NS mergers and the accretion-induced collapse (AIC) of white dwarfs$^{24,25}$ can form massive, rotationally supported NSs that rapidly collapse to black holes in sufficiently sparse environments for FRB radiation to escape$^{11}$. 

Predictions for the birth rate of stellar-mass compact objects that can produce FRBs are also inconsistent with the FRB volumetric rate, if FRB sources do not repeat. Core-collapse supernovae ($\sim10^{5}$\,Gpc$^{-3}$\,yr$^{-1}$) [26] are likely the most common channel for compact-object production, and are consistent with the FRB volumetric-rate limit for low ${\rm DM}_{\rm host}$. However, their young stellar progenitors are commonly associated with star-forming regions of galaxies$^{27}$, and their compact-object remnants would need to migrate into less dense regions and age prior to FRB production to avoid tension with the FRB volumetric rate due to a high ${\rm DM}_{\rm host}$ [16]. The magnetar birth rate is at most half the core-collapse rate$^{28}$, and active magnetars in the Milky Way are only found close to the Galactic plane. Other channels for compact-object production (e.g., NS-NS and WD-WD mergers, AIC of WDs) have even lower volumetric rates. 

If FRBs are associated with stellar-mass compact objects produced through the standard astrophysical channels summarised in Figure~2, the high FRB volumetric rate implies that the majority of FRB sources  repeat. Most FRB progenitor channels require specially configured compact objects, such as young magnetars and NSs in unusual environments$^{14}$. The precise repetition rates of individual FRB sources will depend on the birth rates of the specific progenitor objects, and the lifetimes over which FRBs can be emitted. Recent analyses of the population of FRB sources similar to the repeating FRB\,121102 suggest that this object is atypical of the so far non-repeating FRB population$^{29,30}$. However, for example, even an FRB production rate of $\sim10^{2}$ events per Hubble time per source would be sufficient to resolve the present tension with many of the models discussed here. Ongoing observations with CHIME will provide unprecedented sensitivity to the range of possible FRB repetition rates, together with a larger sample of low-${\rm DM}_{\rm X}$ FRBs to refine the present analysis. The localization of a large sample of FRBs to regions within their host galaxies will enable quantitative tests of population synthesis models for astrophysical FRB progenitors, and estimates of the progenitor ages that will in turn better quantify the expected repeat rates. If the sample completeness can be quantified, FRB distance measurements based on observations of host galaxies will enable a direct measurement of the FRB volumetric rate. 

\vspace{1cm}

{\bf References}

\begin{enumerate}
    
    \item Thornton, D. {\em et al.} A Population of Fast Radio Bursts at Cosmological Distances. {\em Nature} {\bf 341,} 53-56 (2013).
    \item Ravi, V. The observed properties of fast radio bursts. {\em Mon. Not. R. Astron. Soc.} {\bf 482,} 1966-1978 (2019).
    \item Cordes, J. M. \& Lazio, T. J. W. NE2001. I. A New Model for the Galactic Distribution of Free Electrons and its Fluctuations. Preprint at \\ http://arxiv.org/abs/astroph/0207156 (2002).
    \item Shull, J. M. \& Danforth, C. W. The Dispersion of Fast Radio Bursts from a Structured Intergalactic Medium at Redshifts $z < 1.5$. {\em Astrophys. J. Lett.} {\bf 852,} L11 (2018).
    \item Prochaska, J. X. \& Zheng, Y. Probing Galactic haloes with fast radio bursts. {\em Mon. Not. R. Astron. Soc.} {\bf 485,} 648-665 (2019).
    \item Petroff, E. {\em et al.} FRBCAT: The Fast Radio Burst Catalogue. {\em P. Astron. Soc. Aust.} {\bf 33,} e045 (2016).
    \item Spitler, L. {\em et al.} A repeating fast radio burst. {\em Nature} {\bf 531,} 202-205 (2016). 
    \item Amiri, M. {\rm et al.} A second source of repeating fast radio bursts. {\em Nature} {\bf 566,} 235-238 (2019).
    \item Amiri, M. {\em et al.} Observations of fast radio bursts at frequencies down to 400 megahertz. {\em Nature} {\bf 566,} 230-234 (2019).
    \item Shannon, R. M. {\em et al.} The dispersion-brightness relation for fast radio bursts from a wide-field survey. {\em Nature} {\bf 562,} 386-390 (2018).
    \item Kulkarni, S. R., Ofek, E. O., Neill, J. D., Zheng, Z. \& Juric, M. Giant Sparks at Cosmological Distances? {\em Astrophys. J.} {\bf 797,} 70 (2014).
    \item Yang, Y.-P. \& Zhang, B. Extracting host galaxy dispersion measure and constraining cosmological parameters using fast radio burst data. {\em Astrophys. J. Lett.} {\bf 830,} L31 (2016).
    \item Gehrels, N. Confidence limits for small numbers of events in astrophysical data. {\em Astrophys. J.} {\bf 303,} 336-346 (1986).
    \item Platts, E. {\em et al.}, A Living Theory Catalogue for Fast Radio Bursts. Preprint at https://arxiv.org/abs/1810.05836 (2018).
    \item Xu, J. \& Han, J. L. Extragalactic dispersion measures of fast radio bursts. {\em Res. Astron. Astrophys.} {\bf 15,} 1629 (2015).
    \item Walker, C. R. H., Ma, Y.-Z. \& Breton, R. P. Constraining Redshifts of Unlocalised Fast Radio Bursts. Preprint at https://arxiv.org/abs/1804.01548 (2018). 
    \item Li, D., Yalinewich, A. \& Breysse, P. C. Statistical inference of the distance to ASKAP FRBs. Preprint at https://arxiv.org/abs/1902.10120 (2019). 
    \item Kashiyama, K., Ioka, K. \& Meszaros, P. Cosmological Fast Radio Bursts from Binary White Dwarf Mergers. {\em Astrophys. J. Lett.} {\bf 776,} L39 (2013).
    \item Totani, T. Cosmological Fast Radio Bursts from Binary Neutron Star Mergers. {\em Publ. Astron. Soc. Jpn.} {\bf 65,} L12 (2013). 
    \item Falcke, H. \& Rezzolla, L. Fast radio bursts: the last sign of supramassive neutron stars. {\em Astron. Astrophys.} {\bf 562,} A137 (2013).
    \item Abbott, B. P. {\em et al.} GW170817: Observation of Gravitational Waves from a Binary Neutron Star Inspiral. {\em Phys. Rev. Lett.} {\bf 119,} 161101 (2017).
    \item Ruiter, A. J., Belczynski, K., Benacquista, M. \& Holley-Bockelmann, K. The Contribution of Halo White Dwarf Binaries to the Laser Interferometer Space Antenna Signal. {\em Astrophys. J.} {\bf 693,} 383-387 (2009).
    \item Badenes, C. \& Maoz, D. The Merger Rate of Binary White Dwarfs in the Galactic Disk. {\em Astrophys. J. Lett.} {\bf 749,} L11 (2012). 
    \item Moriya, T. J. Radio Transients from Accretion-induced Collapse of White Dwarfs. {\em Astrophys. J. Lett.} {\bf 830,} L38 (2016). 
    \item Ruiter, A. J. et al. On the formation of neutron stars via accretion-induced collapse in binaries. {\em Mon. Not. R. Astron. Soc.} {\bf 484,} 698-711 (2019). 
    \item Taylor, M. {\em et al.} The Core Collapse Supernova Rate from the SDSS-II Supernova Survey. {\em Astrophys. J.} {\bf 792,} 135 (2014).
    \item Kelly, P. L. \& Kirshner, R. P. Core-collapse Supernovae and Host Galaxy Stellar Populations. {\em Astrophys. J.} {\bf 759,} 107 (2012).
    \item Keane, E. F. \& Kramer, M. On the birthrates of Galactic neutron stars. {\em Mon. Not. R. Astron. Soc.} {\bf 391,} 2009-2016 (2008). 
    \item Caleb, M., Stappers, B. W., Rajwade, K. \& Flynn, C. Are all fast radio bursts repeating sources? {\em Mon. Not. R. Astron. Soc.} {\bf 485,} 5500-5508 (2019).
    \item James, C. W. Limit on the population of repeating fast radio bursts from the ASKAP/CRAFT lat50 survey. Preprint at https://arxiv.org/abs/1902.04932 (2019). 
    \item Li, W. {\em et al.} Nearby supernova rates from the Lick Observatory Supernova Search - III. The rate-size relation, and the rates as a function of galaxy Hubble type and colour. {\em Mon. Not. R. Astron. Soc.} {\bf 412,} 1473-1507 (2011).
    \item Ofek. E. O. Soft Gamma-Ray Repeaters in Nearby Galaxies: Rate, Luminosity Function, and Fraction among Short Gamma-Ray Bursts. {\em Astrophys. J.} {\bf 659,} 339-346 (2007).
    
\end{enumerate}

\vspace{1cm}

\noindent Correspondence and requests for materials should be addressed to V. Ravi. 

\vspace{0.5cm}

\noindent {\bf Acknowledgements.} I thank E. Thomas, C. Bochenek and J.-P. Macquart for useful discussions. This work made use of the \texttt{astropy} (http://www.astropy.org) python package. V.R. is supported by a Clay Postdoctoral Fellowship of the Smithsonian Astrophysical Observatory. 

\vspace{0.5cm}

\noindent {\bf Author contributions} V.R. was the sole contributor to this submission.  

\vspace{0.5cm}

\noindent {\bf Competing interests.} The author declares that they have no competing financial interests. 

\vspace{0.5cm}

\noindent {\bf Data availability statement.} The datasets analysed during the current study are available from the FRB Catalogue, http://frbcat.org/. 

\vspace{0.5cm}

\noindent {\bf Code availability statement.} Custom code used in this study is available at \\ https://github.com/VR-DSA/frb\_rate. 

\clearpage

\begin{figure}
    \centering
    \includegraphics[width=0.9\textwidth]{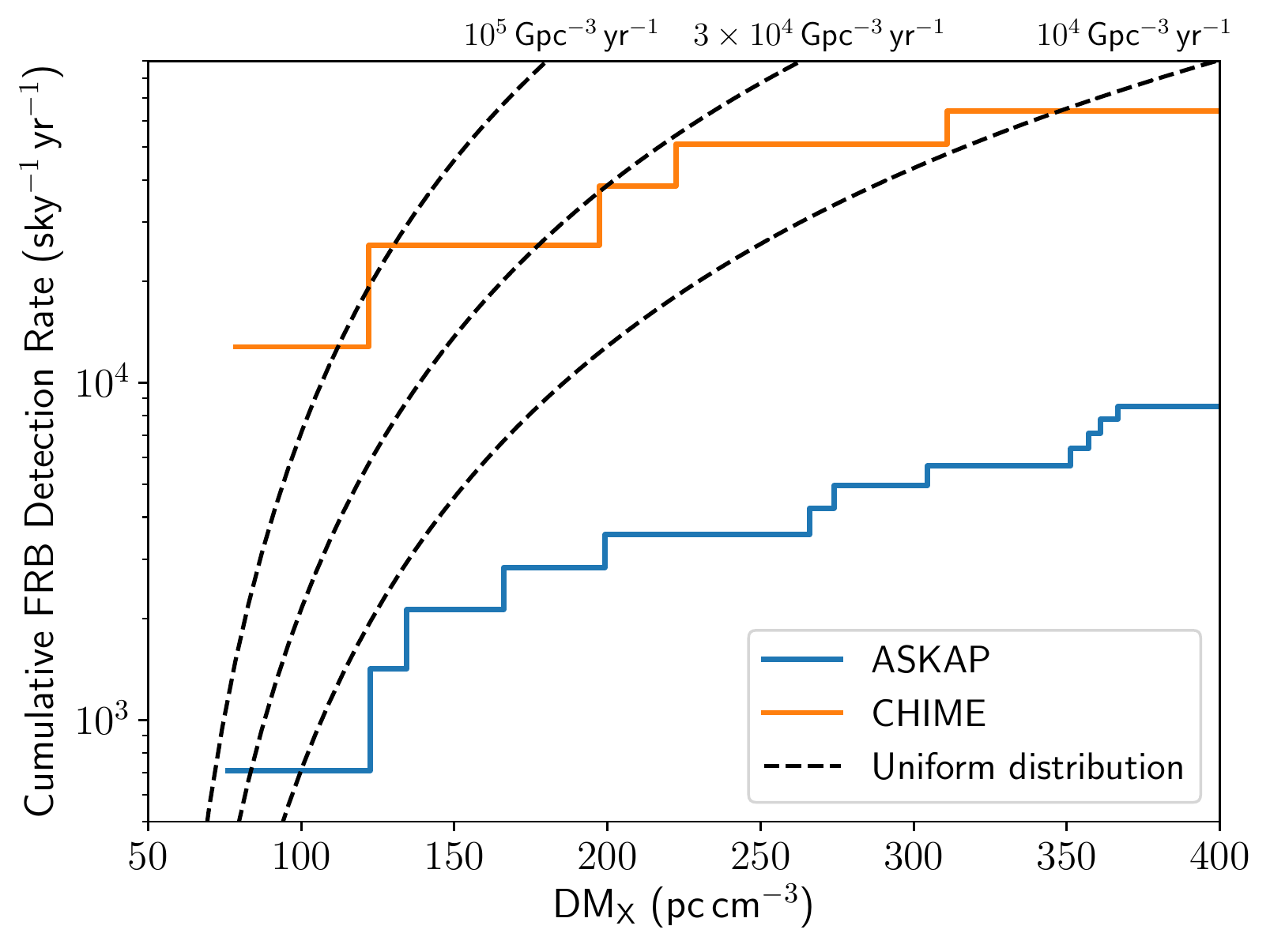}
     \label{fig:1}
    \caption*{{\bf Figure 1: Comparison of the detection rates of ASKAP and CHIME for FRBs within different extragalactic DMs.} The detection rates were calculated assuming survey extents of $1.41\times10^{-3}$\,sky-years for ASKAP$^{10}$ and $7.82\times10^{-5}$\,sky-years for CHIME$^{9}$. At each extragalactic DM (${\rm DM}_{\rm X}$), the detection rate was calculated by dividing the number of FRBs at or below that ${\rm DM}_{\rm X}$ by the survey extent. The dashed lines show curves of predicted cumulative FRB detection rates assuming various comoving volumetric rates (noted at the top of the figure) with no observational incompleteness, ${\rm DM}_{\rm MWhalo} = 50$\,pc\,cm$^{-3}$, and ${\rm DM}_{\rm host} = 0$\,pc\,cm$^{-3}$ (see Methods).}
\end{figure}

\clearpage

\begin{figure}
    \centering
    \includegraphics[width=0.9\textwidth]{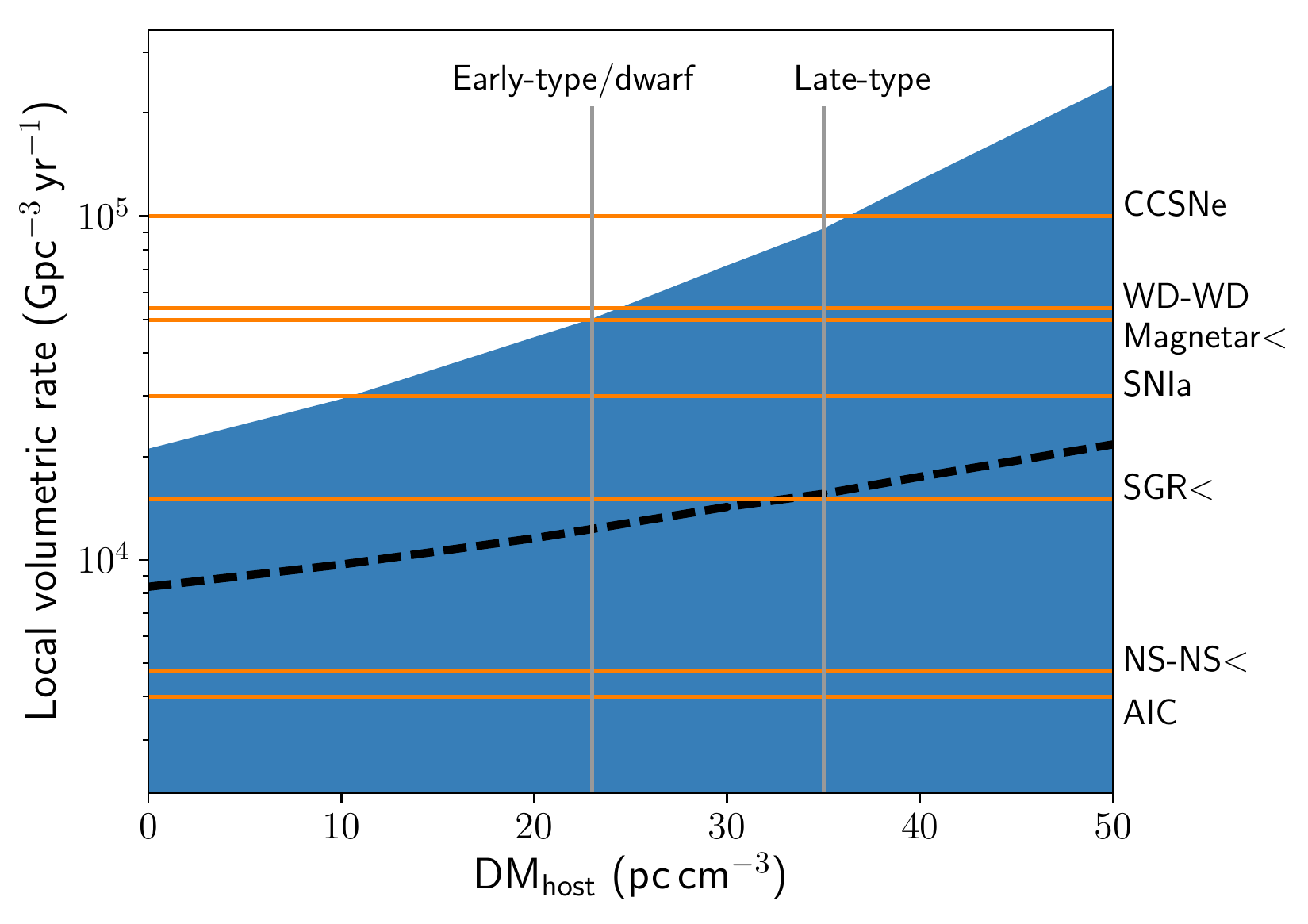}
     \label{fig:2}
    \caption*{{\bf Figure 2: Lower limits on the FRB volumetric rate for different characteristic host-galaxy DMs.} Rates within the blue shaded region are excluded with 90\% confidence given a $d_{\rm limit}$ corresponding to two CHIME FRBs, and the black dashed line indicates the 90\%-confidence limit given a $d_{\rm limit}$ corresponding to three CHIME FRBs. The limits were calculated using FRBs observed during the CHIME pre-commissioning phase, with a detection threshold of approximately 1\,Jy\,ms for millisecond-duration FRBs$^{9}$, and assuming Poisson statistics for the occurrence of so far non-repeating FRBs. The horizontal orange lines indicate estimates of the occurrence rates of candidate cataclysmic FRB progenitors, and the birth rates of candidate compact-object FRB sources. The former includes WD-WD mergers$^{23}$, of which Type Ia supernovae (SNIa) may be a special case$^{31}$, NS-NS mergers (a recent observational upper limit is indicated)$^{21}$, and AIC of WDs$^{25}$. The latter includes NSs produced in core-collapse supernovae (CCSNe), and neutron stars born as magnetars (an upper limit on the highly uncertain rate is shown)$^{28}$ and through the WD-WD, NS-NS and AIC channels. I also show an upper limit on the rate of SGR giant flares$^{32}$, which are a popular scenario for FRB production$^{14}$. The vertical grey lines show the characteristic ${\rm DM}_{\rm host}$ values for orientation-averaged early-type and dwarf galaxies, and for face-on late-type galaxies.}
\end{figure}

\clearpage

\begin{center}
    {\Large \bf Methods}
\end{center}
\vspace{0.5cm}

The choice of the lowest possible $d_{\rm limit}$ values to contain either two or three FRBs does not upwardly bias the lower limit on the FRB rate. I demonstrate this through a simulation. Consider FRBs that are uniformly distributed in space, with a rate of one event per unit time per unit volume. I simulated $10^{4}$ samples of $10^{3}$ events, where in each sample the actual number of events was drawn from a Poisson distribution with a rate parameter of $10^{3}$. I then recorded the distances to the nearest three FRBs in each sample ($d_{1}$, $d_{2}$, and $d_{3}$ respectively). 90\%-confidence lower limits on the volumetric rate were calculated as in the main text$^{13}$: $R_{1} = 0.105\times[(4/3)\pi d_{1}^{3}]^{-1}$, $R_{2} = 0.532\times[(4/3)\pi d_{2}^{3}]^{-1}$ and $R_{3} = 1.102\times[(4/3)\pi d_{3}^{3}]^{-1}$ for each sample. Histograms of $R_{1}$, $R_{2}$ and $R_{3}$ are shown in Supplementary Figure 2, and the 90th percentiles of the samples are indicated. In each case, the 90th percentiles are very close to the true rate of unity. This exercise was also carried out for $10^{4}$ samples of Poisson-distributed events with an expectation of 12, with identical results. 

The use of a low $d_{\rm limit}$ is also motivated by the observations. Figure~1 of the main text shows the empirical cumulative distributions of FRB ${\rm DM}_{\rm X}$ values from ASKAP and CHIME, quantified as all-sky detection rates using the survey extents. I also show curves of predicted FRB all-sky detection rates assuming various comoving volumetric rates and no observational incompleteness, ${\rm DM}_{\rm MWhalo} = 50$\,pc\,cm$^{-3}$, ${\rm DM}_{\rm host} = 0$\,pc\,cm$^{-3}$, and assuming the same DM-distance relation (with no scatter) as in the main text. Cosmological time-dilation effects are included. The shallowness of the CHIME (and ASKAP) empirical distribution of ${\rm DM}_{\rm X}$ values in comparison with the predicted curves suggests incompleteness that increases with ${\rm DM}_{\rm X}$. A similar argument was made in Ref. [33] (e.g., their Figure 1). This effect is also evidenced by the decrease in the lower limits on the FRB rate, $R$, for increasing values of $d_{\rm limit}$. For example, the 90\% confidence lower limits on $R$ for $d_{\rm limit}$ corresponding to four and five FRBs are $R>8.9\times10^{3}$\,Gpc$^{-3}$\,yr$^{-1}$ $R>4.8\times10^{3}$\,Gpc$^{-3}$\,yr$^{-1}$ respectively, for ${\rm DM}_{\rm host}=0$\,pc\,cm$^{-3}$. These arguments only hold as long as ${\rm DM}_{\rm X}$ values are typically dominated by ${\rm DM}_{\rm IGM}$. However, if this were not the case, the conclusions reached in the main text would only be strengthened. 

The repeating CHIME FRB 180814.J0422$+$73 was not included in the present analysis despite its low ${\rm DM}_{\rm X}=102.4$\,pc\,cm$^{-3}$ [8]. This is partly because of the evidence that several properties of the first repeating FRB\,121102, some of which (e.g., the observed repetition, and the burst time-frequency structure) are inconsistent with the population of FRBs that have not been observed to repeat$^{2,29,30}$. Additionally, although the ${\rm DM}_{\rm X}$ of FRB\,180814.J0422$+$73 is lower than that of FRB\,180810.J1159$+$83, the conclusions reached in the main text are independent of its inclusion in the analysis. 

The results presented in the main text are robust to the choice of model for the DM contribution from the Milky Way ionised interstellar medium (${\rm DM}_{\rm MW}$). For example, if the YMW17$^{34}$ model is used instead of the NE2001 model, the lower limit on the FRB rate, $R$, derived for $d_{\rm limit}$ corresponding to two FRBs and ${\rm DM}_{\rm host}=0$\,pc\,cm$^{-3}$ changes from $R>2.1\times10^{4}$\,Gpc$^{-3}$\,yr$^{-1}$ to $R>1.8\times10^{4}$\,Gpc$^{-3}$\,yr$^{-1}$.

\vspace{0.5cm}

\begin{enumerate}
\item Nicholl, M. {\em et al.} Empirical Constraints on the Origin of Fast Radio Bursts: Volumetric Rates and Host Galaxy Demographics as a Test of Millisecond Magnetar Connection. {\em Astrophys. J.} {\bf 843,} 84 (2017).
\item Yao, J. M., Manchester, R. N. \& Wang, N. A New Electron-density Model for Estimation of Pulsar and FRB Distances. {\em Astrophys. J.} {\bf 835,} 29 (2017).
\end{enumerate}

\vspace{0.5cm}

\begin{figure}[h]
    \centering
    \includegraphics[width=0.9\textwidth]{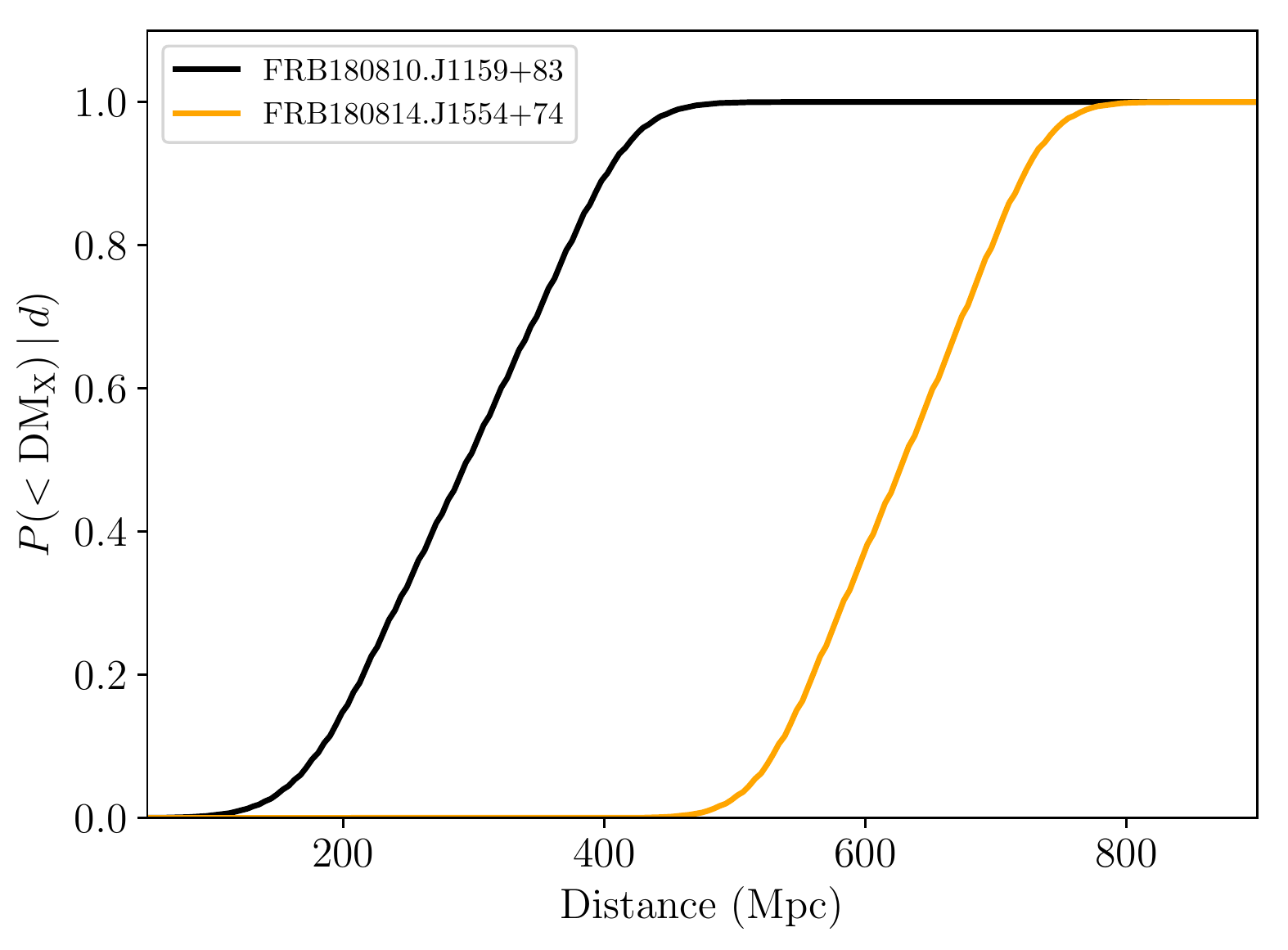}
    \caption*{{\bf Supplementary Figure 1: Probabilities of FRBs 180810.J1159+83 and 180814.J1554+74 having DMs consistent with originating within various distances.} The curves show evaluations of the probability distribution $P(<{\rm DM}_{\rm X}\, | \, d)$ referred to in the main text.}
    \label{fig:3}
\end{figure}

\clearpage

\begin{figure}
    \centering
    \includegraphics[width=0.9\textwidth]{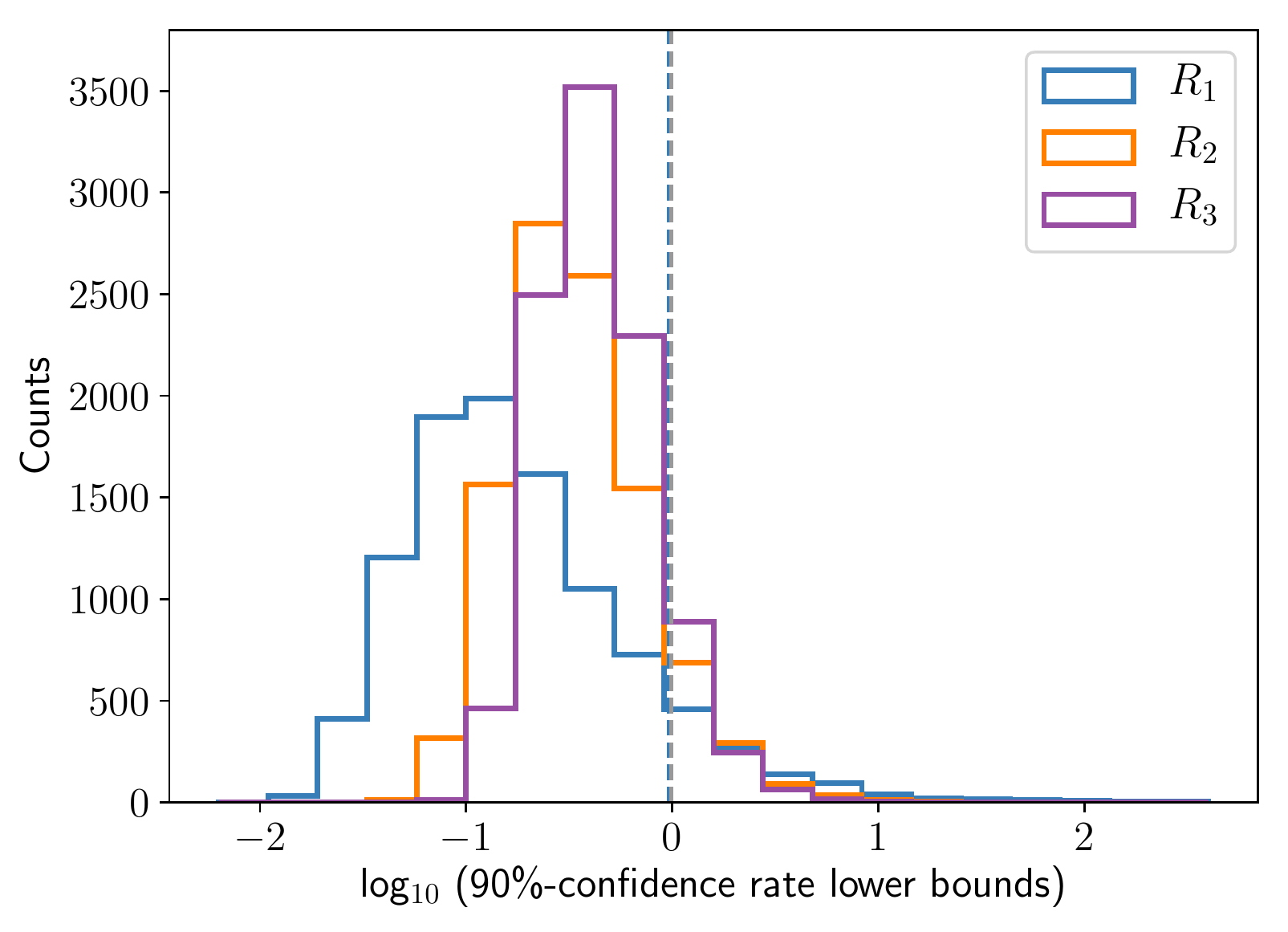}
    \caption*{{\bf Supplementary Figure 2: Histograms of 90\%-confidence lower-limit estimates of the occurrence rate of simulated FRB samples (see Methods).} The true rate in each case was unity. The overlapping vertical dashed lines indicate the 90th percentiles of the distributions.}
    \label{fig:4}
\end{figure}

\end{document}